\documentclass[aps, pre, onecolumn, superscriptaddress,longbibliography]{revtex4-1} 
\usepackage[utf8]{inputenc}
\usepackage{graphicx}
\usepackage{amsmath}
\usepackage{siunitx}
\usepackage{xcolor}

\newcommand{\makered}[1]{\textcolor{black}{#1}}

\begin{document}

\title{Statistics of rigid fibers in strongly sheared turbulence} 

\author{Dennis~Bakhuis}
\affiliation{Physics of Fluids Group and Max Planck
Center Twente for Complex Fluid Dynamics, MESA+ Institute, University of
Twente, Enschede, The Netherlands}

\author{Varghese~Mathai}
\affiliation{Physics of Fluids Group and Max Planck
Center Twente for Complex Fluid Dynamics, MESA+ Institute, University of
Twente, Enschede, The Netherlands} \affiliation{School of Engineering, Brown
University, Providence, RI 02912, USA}

\author{Ruben~A.~Verschoof}
\affiliation{Physics of Fluids Group and Max
Planck Center Twente for Complex Fluid Dynamics, MESA+ Institute, University
of Twente, Enschede, The Netherlands}

\author{Rodrigo~Ezeta}
\affiliation{Physics of Fluids Group and Max Planck
Center Twente for Complex Fluid Dynamics, MESA+ Institute, University of
Twente, Enschede, The Netherlands}

\author{Detlef~Lohse}
\affiliation{Physics of Fluids Group and Max Planck
Center Twente for Complex Fluid Dynamics, MESA+ Institute, University of
Twente, Enschede, The Netherlands}
\affiliation{Max Planck Institute for Dynamics and Self-Organization, Am Fa{\ss}berg 17, G\"ottingen, Germany}

\author{Sander~G.~Huisman}
\affiliation{Physics of Fluids Group and Max Planck
Center Twente for Complex Fluid Dynamics, MESA+ Institute, University of
Twente, Enschede, The Netherlands}

\author{Chao~Sun}
\email{chaosun@tsinghua.edu.cn}
\affiliation{Center for Combustion Energy, Key Laboratory
for Thermal Science and Power Engineering of Ministry of Education, Department
of Energy and Power Engineering, Tsinghua University, Beijing, China}
\affiliation{Physics of Fluids Group and Max Planck Center Twente for Complex
Fluid Dynamics, MESA+ Institute, University of Twente, Enschede, The
Netherlands}

\date{\today}

\begin{abstract}
Practically all flows are turbulent in nature and contain some kind of
irregularly-shaped particles, e.g. dirt, pollen, or life forms such as
bacteria or insects.
The effect of the particles on such flows and vice-versa are highly
non-trivial and are not completely understood, particularly when the particles are finite-sized.
Here we report an experimental study of millimetric fibers in a strongly
sheared turbulent flow.
\makered{%
We find that the fibers show a 
preferred orientation of $-0.38\pi \pm 0.05\pi$ (\SI[separate-uncertainty =
true,multi-part-units=single]{-68(9)}{\degree}) with respect to the mean flow direction in high-Reynolds number Taylor--Couette turbulence,
for all studied Reynolds numbers, fiber concentrations, and locations.
}%
Despite the finite-size of the anisotropic particles, we can explain the
preferential alignment by using Jefferey's equation, which
provides evidence of the benefit of a simplified point-particle approach.
Furthermore, the fiber angular velocity \makered{is strongly} intermittent, 
again indicative of point-particle-like behavior in turbulence.
Thus large anisotropic particles still can retain signatures of the local flow
despite classical spatial and temporal filtering effects.
\end{abstract}
{\let\clearpage\relax\maketitle}
Control and prediction of flows containing anisotropic particles are important
for many industrial settings. For example, in the paper production process,
the alignment of the fibers of the pulp determines the mechanical strength of
the paper \citep{Lundell2011}. In nature, one objective is on flow prediction,
e.g. the dispersion of pollen and seeds \citep{Sabban2017} or sediment
transport in rivers \citep{Vercruysse2017,Lopez2017}. The addition of fibers to the flow
can have significant consequences on the rheology of the suspensions
\citep{Butler2018, Daghooghi2015}.
\makered{
In homogeneous and isotropic turbulence, rod-like fibers
can become preferentially aligned with the vorticity vector
\citep{Wilkinson2009, Parsa2011, Pumir2011, Parsa2012, Voth2017}.
Both, the fibers and the vorticity, are aligned with the largest Lagrangian stretching vector \citep{Ni2014}.
}
When the
fibers behave as tracers, their orientations become correlated with the local
velocity gradients in the flow, and this alignment strongly depends on the
fiber shape \citep{Parsa2012}. In the case of prolate spheroids, the
orientation vector is likely to align with the axis of symmetry of the
flow~\citep{Vincenzi2013}.  In comparison, the behavior of fibers in viscous
shear flows can be noticeably different. Here, the fiber orientation is a
result of the competition between alignment by mean velocity gradients and
randomization by fluctuating velocity gradients \cite{Voth2017}.  This can
lead to either an alignment parallel to the flow direction \citep{Jeffery1922,
Butler1999} or at an angle with the wall \cite{zhang2001, mortensen2008,
Marchioli2010, Marchioli2013, Challabotla2015c, Zhao2015b, Zhao2015}.
However, most of the studies in shear flows have been done by numerical
simulations, addressing the simplified case of inertial point-like fibers
without gravity.  Often a point-particle approach is used, which is considered
to be limited in its applicability to small sub-Kolmogorov scale
\cite{Kolmogorov1941, Kolmogorov1941b} particles, and they have a negligible
particle Reynolds number \cite{Maxey1983}.\\
\indent In most practical situations, however, the suspended particles are not small,
and they have a finite Reynolds number.  Fully resolved numerical simulations,
addressing the effect of fibers in turbulent channel flows showed
that finite size effects lead to fiber--turbulence
interactions that are significantly different from those of point-like
particles\cite{DoQuang2014}. This can lead to an increased dissipation near the particle, and
decreased dissipation in its wake.  In such situations, no analytic
expressions are available for the forces and torques acting on the particles.
In general, it is considered that such finite-sized particles  filter out the
spatial and temporal flow fluctuations~\cite{Toschi2009, Calzavarini2009,
Bec2010, Bellani2012, Bellani2012b, Mathai2015, Almeras2017, Bakhuis2018,
Mathai2018}, and hence do not actively respond to the local gradients in the
flow. Few experiments have explored this regime of finite sized rod-like
fibers in sheared turbulence.\\
\begin{figure*}[htp] 
\centering
\includegraphics{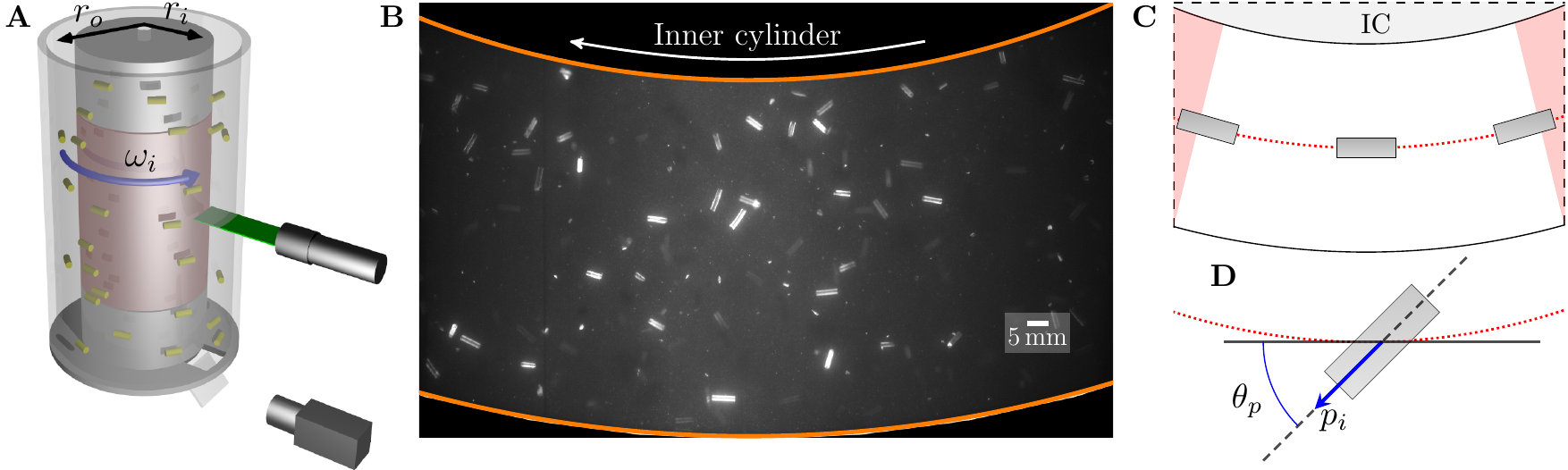}
\caption{(A) Schematic of the experimental apparatus (not to scale). The flow
is confined between two concentric independently rotating cylinders with radii
$r_i$ and $r_o$. Only the inner cylinder~(IC) rotates with an angular velocity
$\omega_i$. A mirror and a window in the bottom plate provide optical access
to the $r$-$\theta$ plane for a high-speed camera. (B) A typical still image
with the inner and outer cylinder highlighted in orange.  \makered{The fibers (aspect ratio $\Lambda =5.3$) are clearly visible as white rods.} $\text{Re}_i = \num{1.7e5}$ and $\alpha = \SI{0.05}{\percent}$. (C) Schematic of the $r$-$\theta$ plane. The orientation of the particle, $\theta_p$, is zero when it is aligned with the IC. Fibers with their center in the red areas are removed from all statistics. (D) Definition of the orientation, $\theta_p$, and the orientation vector, $p_i$, of a fiber. $\theta_p$ is measured with respect to the azimuthal direction and is defined positive in the counter-clockwise direction. \makered{See Movies S1 and S2 in the supplemental material for typical recordings showing the movement and tracking of the fibers \citep{Video1,Video2}.}}
\label{fig:setup}
\end{figure*}
%
\indent In this letter we probe the dynamics of a suspension of millimetric
rod-like fibers in a strongly turbulent Taylor--Couette (TC) flow~(see
Fig.~\ref{fig:setup}AB). The reason for choosing this geometry is at least
three-fold: (i) it is a closed geometry, allowing for direct relationships
between local and global quantities \cite{Eckhardt2007}, (ii) there are no
spatial transients, \textit{i.e.}, the turbulence intensity does not depend on
the streamwise position as it does in channels and pipes, and (iii) it allows
for high Reynolds numbers in a limited space \cite{Grossmann2016}. %
All experiments are conducted in the Twente Turbulent Taylor--Couette (T$^3$C)
facility \cite{vanGils2011}, which confines the flow between two concentric
cylinders (see \ref{fig:setup}A).
The inner and outer cylinders radii are $r_i=\SI{0.2000}{\metre}$ and
$r_o=\SI{0.2794}{\metre}$, respectively, giving a radius ratio of
$\eta=r_i/r_o=0.716$ and a gap width $d=\SI{79.4}{\milli\metre}$. The height
of the system is $L=\SI{0.927}{\metre}$, which results in an aspect ratio of
$\Gamma=L/d=11.7$. We rotate the inner cylinder (IC) with angular velocity
$\omega_i$ while the outer cylinder (OC) is kept at rest. \makered{The flow is seeded with rigid fibers of length 
$\ell = \SI[separate-uncertainty =
true,multi-part-units=single]{5.22(7)}{\milli\metre}$,
cut from a PMMA optical fiber of diameter 
$d_p =\SI[separate-uncertainty = true,multi-part-units=single]{0.99(1)}{\milli\metre}$
(aspect ratio $\Lambda=\ell/d_p = 5.3\pm0.1$)}.
The 2D projection of the orientation angle on the
radial-azimuthal plane, $\theta_p$, is defined to be zero when the fiber is
aligned with the IC (Fig.~\ref{fig:setup}C) and positive values
are in the counter-clockwise direction (Fig.~\ref{fig:setup}D).
\makered{To minimize density effects, glycerol and water are mixed 1:1, giving a density ratio of $\rho_p/\rho_\text{fluid}=\SI{1210}{\kilo\gram\per\metre\cubed}/\SI{1140}{\kilo\gram\per\metre\cubed}=1.06$.}
The dominant velocity is in the azimuthal
direction. Velocities in the axial and radial directions are due to secondary
flows and are approximately \SI{5}{\percent} of the azimuthal velocity. While
the particles are free to rotate in all directions, the largest velocity
gradient is in the radial direction, resulting in a rotation in the axial
direction.%
\begin{figure}[htp] 
\centering
\includegraphics[scale=1.2]{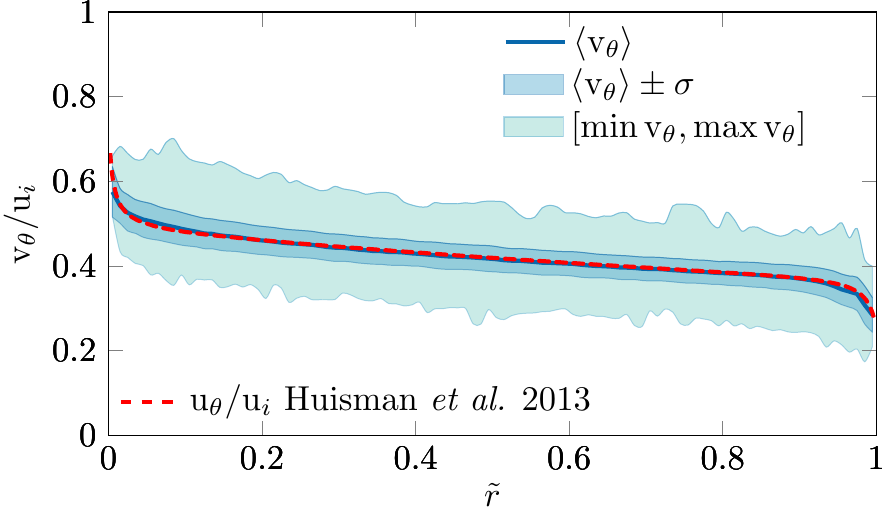}
\caption{Fiber velocity as a function of the dimensionless radius
$\tilde{r}=(r-r_i)/d$ for $\text{Re}_i=\num{1.7e5}$, $\alpha =
\SI{0.05}{\percent}$, and $\tilde{z} = z/L = 0.24$. All velocities are
normalized using the velocity of the IC $u_i$. For comparison, the
azimuthal flow profile is included as a red dashed line.  We find that the
azimuthal velocity of the fibers is very close to the velocity of the flow.}
\label{fig:fibervelocity}
\end{figure}
For the flow under consideration, the control parameters
are the Reynolds number $\text{Re}_i=\omega_i r_i (r_o - r_i)/\nu$ and the
volume fraction of the fibers $\alpha$. Here $\nu$ is the the kinematic viscosity.  $\text{Re}_i$ is varied by changing $\omega_i$, resulting in a $\text{Re}_i$
range from $8.3 \times 10^4$ to $2.5\times10^5$ which lies in the so-called
ultimate regime \cite{Lathrop1992, vanGils2011, Huisman2012b, Ostilla-Monico2014} of
turbulent Taylor--Couette flow, where both the bulk and boundary layers
are turbulent. From the volume fraction of fibers $\alpha=0.025\%$ to
$\alpha=0.100\%$, the suspensions we study are on the border of dilute and
dense suspensions, which has either two- or four-way coupling
\cite{Elghobashi1994}. 
To capture the orientation and velocities of the fibers, images in the
radial-azimuthal plane are captured using a Photron \texttt{SA-X2} high-speed
camera. \makered{Illumination comes from a Litron \texttt{LDY-303} pulsed laser and sheet optics, creating an approximately \SI{2}{\milli\metre} thick plane.
To improve the contrast, the fibers were coated using rhodamine-B making them fluorescent, and a band-pass filter was used isolate the fiber's signal.}
Fig.~\ref{fig:setup}B shows a typical captured image in which the IC and OC are highlighted \citep{Video1, Video2}.
A total amount of 64 thousand images per case
($\text{Re}_i, \alpha, z/L$) are captured and the position and orientation of each of the fibers are extracted, see figs.~\ref{fig:setup}CD. 
These are then tracked over time,
from which, the velocity, $v_\theta$, and angular velocity,
$\dot{\theta}_p$, can be determined. We find that the fibers distribute nearly
homogeneously in the radial direction of the measurement volume. Moreover, we find that their
azimuthal velocity, normalized using the velocity of the IC,
$u_i$, closely follows the azimuthal velocity profile of the flow, $u_\theta$,
\cite{Huisman2013}, see Fig.~\ref{fig:fibervelocity}. These fibers, therefore,
do not show clustering or relative velocities, which seems surprising
considering their rather large size.  However, the absence of clustering can
be expected, since the fibers are nearly neutrally buoyant
\cite{Calzavarini2008, Calzavarini2009, Fiabane2012}. Yet, this cannot explain
the absence of relative velocities with the flow~(see
Fig.~\ref{fig:fibervelocity}) we observe in our experiment. To explain this
behavior, we calculate the Stokes number $\text{Stk}_K \equiv \tau_v/\tau_K$,
where $\tau_v = \frac{\ell^2}{3 \beta \nu}$ with $\beta =
\frac{3\rho_f}{2\rho_p + \rho_f}$ is the particle response time
\cite{Qureshi2007}, and $\tau_K = \sqrt{\frac{\nu}{\epsilon}}$ is the
Kolmogorov time scale. For our flow conditions we find that $\tau_K =
[\SI{2.2}{\milli\second},\SI{0.5}{\milli\second}]$ and the Kolmogorov length
scale $\eta_K = [\SI{113}{\micro\metre},\SI{52.6}{\micro\metre}]$, where each
two values correspond to our lowest and highest Reynolds numbers $\text{Re}_i
= \num{8.3e4}$ and $\num{2.5e5}$, respectively. These values result then in
$\text{Stk}_K = [110.0, 510.0]$, and size ratios $\ell/\eta_K = [44.0, 95.0]$. This
suggests that the fibers are large and highly inertial, and hence, should
filter out the flow fluctuations \cite{Calzavarini2009, Bec2010,
Marchioli2010, Parsa2011, Marchioli2013, Voth2017}.  We therefore have to
correct our previous Stokes number estimation as the relevant time scale is
not given by $\tau_K$, but rather by the time scale $\tau_\ell$ of turbulent
eddies comparable to the fiber size. $\tau_\ell = \left( \ell^2 / \epsilon
\right)^{1/3}$ \cite{Xu2008}, resulting in $\text{Stk}_\ell \equiv
\tau_v/\tau_\ell = [9.0, 24.0]$. $\text{Stk}_\ell$, though, assumes that the
particles have a tiny Reynolds numbers.  Based on the liquid velocity
fluctuations~\cite{Calzavarini2009}, we calculate the particle Reynolds number
$\text{Re}_p = \sigma(u_\theta) \ell / \nu = \mathcal{O}(10^3)$, with $\sigma$
the standard deviation, which far exceeds the viscous flow limit. Based on
these insights, we use a modified viscous time scale
\cite{Clift1978,Qureshi2007} for the particle $\tau_p$, which also takes into
account the drag coefficient $C_D(\text{Re}_p)$. Remarkably, the resulting
Stokes number $\text{Stk}_p \equiv \tau_p/\tau_\ell = [2.0, 3.0]$, indicating that
the fibers are only slightly inertial, which explains why they follow the flow
field (at their length scale) quite accurately.\\%
\begin{figure}[htp]
\centering
\includegraphics[scale=1.2]{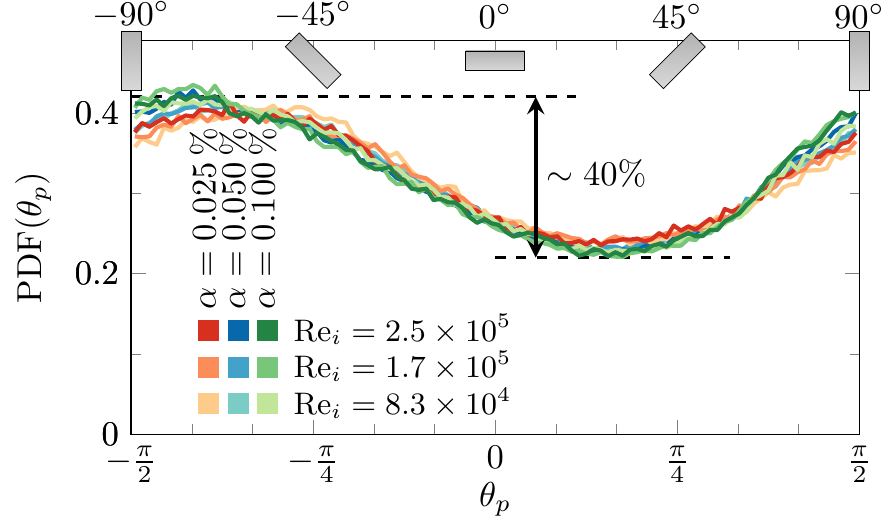}
\caption{PDF of the fiber orientation $\theta_p$ measured at $\tilde{z} =
0.24$. Different $\alpha$ are indicated by different hues and different
$\text{Re}_i$ are shown with different shades. A representation of the fiber
alignment is shown at the top of the figure. \makered{Independent of $\alpha$ and
$\text{Re}_i$ there is a clear preference for an alignment around
$-0.38\pi \pm 0.05\pi$ (\SI[separate-uncertainty =
true,multi-part-units=single]{-68(9)}{\degree}).} A large \SI{40}{\percent} difference between the most and
least probable orientation is observed.}
\label{fig:orientre}
\end{figure}
\begin{figure}[htp]
\centering
\includegraphics[scale=1.2]{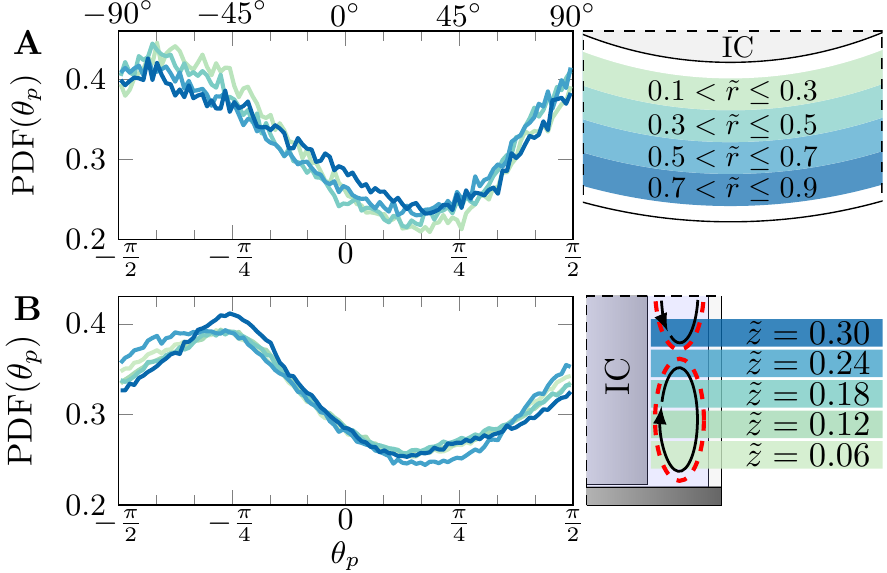}
\caption{(A) PDF of the fiber orientation $\theta_p$ at various radial bins,
indicated by different colors. $\alpha$ is fixed to \SI{0.05}{\percent},
$\text{Re}_i = \num{2.5e5}$, and the measurement is performed at $\tilde{z} =
0.24$. (B) Axial dependence of the PDF of $\theta_p$, indicated by different
colors. For these measurements $\alpha=\SI{0.05}{\percent}$ and
$\text{Re}_i=\num{8.3e4}$. \makered{The diagram on the right indicates the position of the weak vortical structures \cite{Huisman2014,vanderVeen2016}.
The distribution is found to be nearly independent
of radial and axial positions, and all show similar
alignment.}}
\label{fig:orientrz}
\end{figure}

\begin{figure}[htp]
\centering
\includegraphics[scale=1.2]{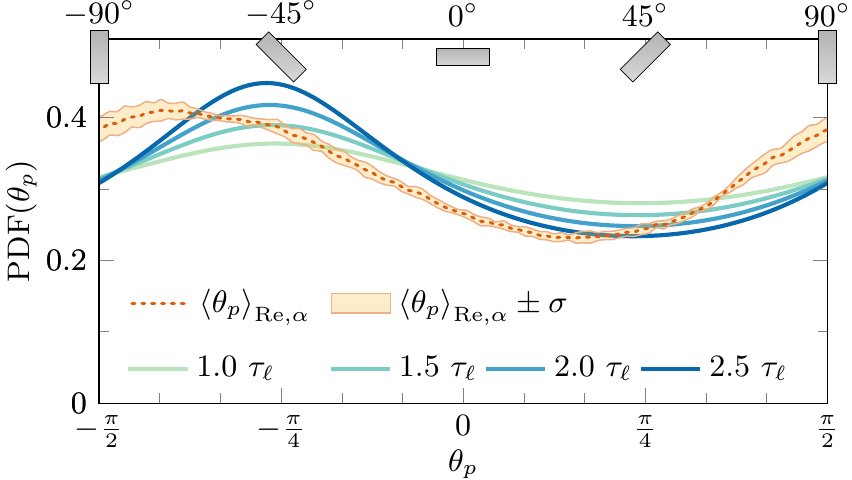}
\caption{Averaged PDF of the experimentally found fiber orientation (dashed)
compared to the alignment found from integrating Jeffery's equations (solid
lines). The legend indicates the integration time scale as multiples of
$\tau_\ell$.}
\label{fig:fiberjeffery}
\end{figure}

\indent Next, we address the orientation statistics of the fibers in the flow.
To check whether or not the fibers show any preferential alignment, we first
look at the probability density function (PDF) of the orientation (see
Fig.~\ref{fig:setup}D for definition) for various $\alpha$ and
$\text{Re}_i$, see Fig.~\ref{fig:orientre}. \makered{We find that for all cases studied, the PDF of the orientation shows a preference for
$\theta_p = -0.38\pi \pm 0.05\pi$} (\SI[separate-uncertainty =
true,multi-part-units=single]{-68(9)}{\degree}). Since Taylor--Couette flow
\cite{Grossmann2016} is known to have (turbulent) Taylor vortices \cite{Huisman2014, vanderVeen2016}
\makered{(Relative positions shown in the right diagram of Fig. \ref{fig:orientrz}B)}, one might expect this preferential alignment to depend on
the axial~($\tilde{z}=z/L$) and radial~($\tilde{r}=(r-r_i)/d$) positions of
the fibers. We therefore provide PDFs conditioned on $\tilde{r}$, and perform
additional measurements at several $\tilde{z}$, see Fig.~\ref{fig:orientrz}.
\makered{Surprisingly, the preferential alignment around $\approx -0.38\pi$
persists throughout the flow. We find nearly identical orientation PDFs for
different $\text{Re}_i$ and $\alpha$, and even at different $\tilde{r}$ and
$\tilde{z}$}.
It is remarkable that a single preferential alignment value exists throughout the flow domain, despite 
the strong flow anisotropies and the finite size of the fibers.
%
\indent In order to understand the preferential alignment of the fibers, we
model their dynamics using a simplified model based on the equations by Jeffery~\cite{Jeffery1922}, derived for ellipsoidal particles in
a viscous fluid in the limit of small Stk and small $\text{Re}_p$. Jeffery's
equations in the non-inertial limit are duplicated here:
\begin{align}
\dot{p}_i = \Omega_{ij} p_j + \frac{\Lambda^2 - 1}{\Lambda^2 + 1} \left(
S_{ij}p_j - p_i p_k S_{kl} p_l \right) \label{eq:jeff}
\end{align}
where $p_i$ is the orientation vector (see Fig.~\ref{fig:setup}D),
$\Omega_{ij}$ is the vorticity tensor
$\Omega_{ij}= \frac 12 \left(\frac{\partial u_i}{\partial x_j} - \frac{\partial u_j}{\partial x_i} \right) = -\epsilon_{ijk}\omega_k$,
where $\epsilon_{ijk}$ is the Levi-Civita symbol
in 3D, $\omega_k$ is the vorticity vector, $\Lambda$ is the aspect ratio of
the particle, and $S_{ij}$ is the strain-rate tensor 
$S_{ij}= \frac 12 \left(\frac{\partial u_i}{\partial x_j} + \frac{\partial u_j}{\partial x_i} \right)$.
We model a stochastic mean field process by assuming that the fibers
are initially randomly oriented by the turbulent fluctuations every time
interval $\sim \mathcal{O}(\tau_\ell)$. Next, we model the flow seen by the
fiber as a simple shear flow, with a mean shear rate equal to that in the bulk
of the turbulent Taylor--Couette flow, i.e.
$\dot \gamma = \langle \partial u_\theta / \partial r \rangle_{\tilde{r}\in[0.25,0.75]}$ from
Fig.~\ref{fig:fibervelocity}. We integrate eqs.~\ref{eq:jeff} for all initial
conditions $p_i(t=0) = \sin \left(\frac{\pi  i}{2}-\theta \right)$ for 
$\theta \in [-\pi/2,\pi/2]$ over a variety of time scales $t \in [0,C\tau_\ell]$ where
$C$ is a dimensionless constant of $\mathcal{O}(1)$, to obtain $p_i(t)$ for
every initial condition. These evolutions of $p_i(t)$ are converted to
$\theta(t)$ using the definition given in Fig.~\ref{fig:setup}D, aggregated,
and binned to calculate the PDF of $\theta$, see Fig.~\ref{fig:fiberjeffery}.
\makered{The PDF shape predicted by this simplified model is remarkably similar to our experimental observations.}
\makered{The preferred orientation (peak) calculated from
Jeffery's equation is approximately $-0.27\pi$ ($\approx$\,\SI{50}{\degree}), which is close to the measured value of $-0.38\pi$ ($\approx$\,\SI{68}{\degree}).}
\makered{%
The shift of approximately \SI{15}{\degree} can be explained by a slight inertial effect: as the particle goes around the inner cylinder it continuously adapts to the local orientation of the flow.
We believe that this discrepancy is likely due to a lag towards this preferred alignment.
Also, the turbulent fluctuations, which are not considered in this model, can cause an offset towards the prefered alignment.
}%
The amplitude of the measured PDF is close to the
calculation with the integration time being $2 \tau_\ell$, which measures the
time scale of rotation of a non-inertial fiber in the bulk of the flow. The
non-inertial approach for orientation modeling
is reasonable, since typically the rotational Stokes number 
$\text{Stk}_r \sim \mathcal{O}(0.1 \ \text{Stk}_p)$ for long prolate ellipsoids \cite{Zhao2015}.
Nevertheless, slight differences between the calculation and experimental
results (seen in Fig.~\ref{fig:fiberjeffery}) are expected, since our fibers
are not truly in the $\text{Stk}_r \to 0$ limit~\cite{Challabotla2015b}.\\%
\begin{figure}[htp]
\centering
\includegraphics[scale=1.2]{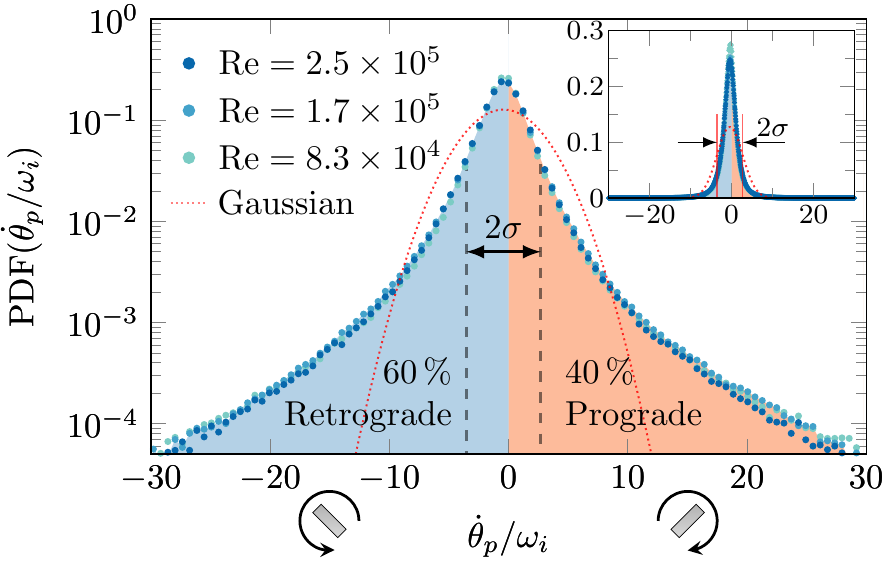}
\caption{PDF of the rotation rate of the fibers for
$\alpha=\SI{0.05}{\percent}$ and $z/L=0.24$. Rotational velocities are
normalized using the angular velocity of the IC. The PDF is
independent of $\text{Re}_i$ and shows a slight preference for retrograde
rotation (blue). Note that the icons holds for CW rotation of the inner
cylinder.
\makered{The mean rotation is $\langle \dot{\theta}_p/\omega_i \rangle \approx -0.42$ with a standard deviation of 
$\sigma(\dot{\theta}_p/\omega_i)=3.13$, which reveals that a large number of 
fibers rotate much faster than the inner cylinder. For comparison, a Gaussian distribution with the same mean and variance is added.}
The skewness and kurtosis are found to lie in
the range $[-0.14, 0.24]$ and $[34,40]$, respectively.
\makered{The inset shows the same data on a linear scale.}%
}%
\label{fig:fiberrotation}%
\end{figure}%
%
\makered{The PDF of the fiber rotation rate at different $\text{Re}_i$ is shown in
Fig.~\ref{fig:fiberrotation} (The inset shows the same data on linear scale).}
The PDF of the normalized rotation rate is found
to be independent of the Reynolds number.
Due to the mean shear in the bulk of the flow, it has a slight preference for retrograde rotation (rotation in the opposite
direction of the IC) with 60\% probability. We notice
that the peak of the PDF is located at $\langle
\dot{\theta_p}/\omega_i \rangle \approx -0.42$, which is comparable to the
mean vorticity in the bulk of the flow.
\makered{The standard deviation is $\sigma(\dot{\theta}_p/\omega_i)=3.13$, which reveals that a large number of fibers rotate much faster than the inner cylinder.}
What is really remarkable is the
strong intermittency of the PDF with tails extending beyond $\pm30\omega_i$,
which occurs despite the large size of the fibers. We find a skewness of the
angular velocity between  $[-0.14, 0.24]$. The kurtosis of the angular
velocity lies in the range $[34,40]$, which is much larger than the kurtosis
of spheres of similar size
ratios~\cite{Zimmermann2011,Mathai2016b,Mathai2018b}. This can be attributed
to the fact that for elongated ellipsoids the rotational inertia is typically
much lower than the rotational inertia of similar-sized
spheres~\cite{Zhao2015}. Furthermore, the length of the fibers is of
$\mathcal{O}(100\eta_K)$, the two ends of the fibers can therefore experience extremely high instantaneous velocity differences due to the intermittent nature of the turbulent velocity fluctuations. These instantaneous velocity differences can create high torques on the fiber, resulting in violent rotational intermittency.\\
\indent To summarize, we report on the statistics of translation and
rotation of finite-sized fibers in an strongly sheared turbulent
flow. The fibers tend to follow the flow almost perfectly,
despite their large size. This adherence to the flow can be explained by
considering the turbulent dynamic time at the scale of the fiber, compensated
by effects of non-linear drag at the finite Reynolds number of the fiber,
yielding a Stokes number estimate that is just above unity.
\makered{For the fiber
orientation statistics, while it was often hypothesized that no systematic
alignment would be possible in such highly turbulent flows with very strong liquid fluctuations, in this canonical TC
flow geometry, we show that fibers do align with an angle of $-0.38\pi \pm 0.05\pi$
(\SI[separate-uncertainty =
true,multi-part-units=single]{-68(9)}{\degree}) with respect to the IC wall.} The difference between the most and least preferred alignment is dramatic, namely \SI{40}{\percent}. This
alignment is \makered{similar for all tested Reynolds numbers~($\text{Re}_i$), fiber
volume fractions~($\alpha$), and spatial coordinates.}
We model the fiber orientation statistics using
Jeffery's equations, which provides a fair estimate of the shape of the
alignment PDFs. Additionally, it is found that the fiber angular velocity
shows extremely high intermittency with instantaneous rotation rates much larger than that of the rotating cylinder. \makered{Thus, in a number of ways,
finite-sized (millimetric) fibers behave remarkably similar to tiny particles
in turbulence~\cite{Voth2017,Toschi2009}, extending the possibilities of the
point-particle approach to model large anisotropic particles in turbulence.}
\begin{acknowledgments}
We thank Eric Climent, Enrico Calzavarini, Vamsi Spandan, Dominik Krug, Jelle Will, Pim Bullee, and Arne te Nijenhuis for various stimulating discussions, and Gert-Wim Bruggert and Martin Bos for technical support. This work was funded by the Natural Science Foundation of China under Grant No. 91852202, the Netherlands Organisation for Scientific Research (NWO) under VIDI grant No. 13477,  STW, FOM, and MCEC.
\end{acknowledgments}

\bibliography{PRF-fibers}

\end{document}